\documentclass{appolb}
\usepackage{epsfig}
\usepackage{amsmath,graphicx}

\begin{document}
%\titlepage                                                    
\title{\vspace*{-6cm}\begin{flushright}                                                    
CERN--PH--TH/2009--067 \\
MAN/HEP/2009/19 \\
HERWIG/09/04 \\
KA-TP-05-2009 \\
TPJU-1/2009\\
MCnet/09/10 \\                                                  
\end{flushright}
\vspace*{2cm}
Non-perturbative effects in the transverse momentum distribution of electroweak bosons at the LHC%
\thanks{
Based on a talk by Andrzej Si\'odmok at the Cracow Epiphany Conference on “Hadron interactions at the dawn 
of the LHC, 5-7 January 2009; dedicated to the memory of Jan Kwiecinski”.}%
}
\author{Andrzej Si\'odmok
\address{ Marian Smoluchowski Institute of Physics, Jagiellonian University,\\
         ul.\ Reymonta 4, 30-059 Cracow, Poland \vspace*{2mm}\\
 LPNHE, Pierre et Marie Curie Universit\'es Paris VI et Paris VII,\\ 
          Tour 33, RdC, 4, pl. Jussieu, 75005 Paris, France
}
\and
Stefan Gieseke
 \address{Institut f\"ur Theoretische Physik,\\ Universit\"at Karlsruhe, 
  76128 Karlsruhe, Germany.}
\and
Michael H.~Seymour
\address{PH Department, CERN, 1211 Geneva 23, Switzerland\\ 
  \emph{and} School of Physics \& Astronomy, University of Manchester, U.K.}
}

%\preprint{ HERWIG/09/03 \\ MCnet/09/09 \\ IPPP/09/41 \\ DCPT/09/82 \\ CP3-09-09 }
\maketitle
\begin{abstract}
The transverse momentum of electroweak bosons in a Drell-Yan process is an important 
quantity for the experimental program at the LHC. The new model of non-perturbative 
gluon emission in an initial state parton shower presented in this note gives a good 
description of this quantity for the data taken in previous experiments over a wide range of CM energy. 
The model's prediction for the transverse momentum distribution of $Z$ bosons for the LHC is 
presented and used for a comparison with other approaches.
\end{abstract}

\newcommand{\hpp}{Herwig++}
\newcommand{\ResBos}{ResBos}
\section{Introduction}
The Drell-Yan process has been widely studied in many past
\cite{Stirling:1993gc} and present \cite{CDFdata,D0data} hadron collider experiments and played a 
significant role in the development of our understanding of QCD and electroweak (EW)
interactions, both from the experimental and theoretical point of view.
Certainly this will also be the case for the LHC experiments, especially because it
will soon become the unique $W$ and $Z$-boson production factory 
which is expected to collect 300 million $W$ and 20 million $Z$ events per year of its operation at energies $\sqrt{s} = 14$ TeV and the luminosity
of $10^{33}\ cm^{-2}s^{-1}$.
Among various distributions of $W$ and $Z$ observables, the transverse momentum spectrum of vector bosons in a Drell-Yan process is a very useful and important quantity for the experimental program at the LHC.
In the case of $W$ production, the uncertainty in the shape of 
the spectrum directly affects the measurement of the $W$ mass \cite{Besson:2008zs} and its mass charge asymmetry $M_{W^+}- M_{W^-}$ \cite{Fayette:2008wt}.
It also helps to understand the signature for Higgs boson
production at either Tevatron or LHC \cite{Balazs:2000sz}.
Although the experiments measure the $Z$ transverse momentum distribution
and use this to infer that of the $W$ boson, the extent to which the effects are
non-universal limits the ultimate accuracy of the measurement, unless
elaborate tricks, as proposed in Ref. \cite{StandardCandle} are used.
For these reasons, it is of utmost importance to predict the $W$ and $Z$ observables 
with as high as possible theoretical precision. 
The sources of uncertainty in the theoretical predictions of observables, such as 
the transverse momentum of electroweak bosons discussed here, 
are of perturbative and non-perturbative origin.
In this short note we will concentate on the modelling of the latter in the framework of 
a backward evolution parton shower approach \cite{Sjostrand:1985xi} which is widely used in general 
purpose Monte Carlo Generators such as Herwig \cite{Corcella:2000bw}, Pythia \cite{Sjostrand:2006za} or Sherpa \cite{Gleisberg:2008ta}.
During the parton shower evolution, which terminates at some scale of 
typical hadron mass, the recoil from the emitted gluons\footnote{Together with other 
backward-evolution steps, such as an incoming sea-quark being evolved back to an incoming 
gluon by emitting a corresponding antiquark.} builds up a transverse momentum for the $W/Z$.
In order to fit existing data, the conventional backward evolution parton shower approach 
needs to be supplemented by the so-called ‘intrinsic’ (or `primordial') transverse momentum 
$k_T$ distribution of partons initiating the shower. The physical motivation behind this 
additional non-perturbative ingredient is the Fermi motion of partons within a hadron.
Therefore, its average value per parton can be estimated based solely on 
the proton size and uncertainty principle to be of the order of $0.3$ -- $0.5$ GeV. 
But the values extracted from data first of all are too large and secondly grow with 
collision energy which cannot be explained by Fermi motion. For 
example, in Herwig++ its value grows from $k_T = 0.9$ GeV, which is needed to describe the 
data taken at the energy $\sqrt{s}=62$ GeV (experiment R209), to $2.1$ GeV which, is needed 
at the Tevatron energies ($\sqrt{s} = 1800$ GeV). This motivated us to propose a model 
for backward evolution in which an additional non-perturbative component at low transverse momentum
provides additional smearing at each step of the evolution. By construction we
expect more non-perturbative smearing for longer parton shower evolution  ladder which might cure the problem
of dependence on centre of mass energy as well as on the size of needed intrinsic smearing, which 
in our studies is kept, according to the Fermi motion argumentation fixed at $0.4$ GeV. 

In the following sections we will first briefly describe the model, then present how it fits the existing data sets, and at the end of this note we will demonstrate the model's predictions for the LHC energies which we use for a 
comparison with other approaches.
\section{Model}
The implementation of transverse momentum production in which non-perturbative smearing takes place throughout the perturbative evolution, 
was achieved by a simple modification to an initial-state parton shower algorithm. The model was implemented in the framework of  \hpp{} \ \cite{herwigman}
in which the Sudakov form factor for backward evolution from some
scale $\tilde q_{\rm max}$ down to $\tilde q$ takes the form
\begin{equation}
  \label{eq:sudakov}
  \Delta(\tilde q; p_{\perp_{\rm max}},p_{\perp_0}) = \exp 
  \left\{-\int_{\tilde q^2}^{\tilde q^2_{\rm max}}
    \frac{d\tilde q'^2}{\tilde q'^2} \int_{z_0}^{z_1} 
    dz \frac{\alpha_s(p_{\perp})}{2\pi} 
    \frac{x'f_b(x', \tilde q'^2)}{xf_a(x, \tilde q'^2)}
  P_{ba}(z, \tilde q'^2)\right\} \ ,
\end{equation}
with $x'=x/z$, for further details cf.\
Ref.~\cite{Gieseke:2003rz}. 

The argument of the strong coupling $\alpha_s$ in
Eq.~(\ref{eq:sudakov}) is the transverse momentum $p_{\perp}$ of an
emission\footnote{In Herwig++,
the argument of $\alpha_s$ is a slightly simplified expression, equal
to the transverse momentum to the required accuracy, but not exactly.
We have tested the implementation of our model with this simplified
expression and the exact expression for transverse momentum, and find
very similar results.  We therefore use the default expression.}.  
The cut-off scale represented by $p_{\perp_0}$ is needed to avoid divergence
of the strong coupling. Below the cut-off scale $\alpha_s$ is equal to zero and consequently
the derivative of the Sudakov form factor is equal to zero which translates to zero
probability of the gluon emission below $p_{\perp_0}$. 
Therefore, the two arguments of the Sudakov form factor, $p_{\perp_{\rm max}}$ and
$p_{\perp_0}$, are not the evolution variables but only explicitly specify
the available phase-space of an emission.

In order to populate the phase-space below  $p_{\perp_0}$ by additional non--perturbative emissions 
we introduce the additional Sudakov form factor $\Delta_{NP}$ such that
\begin{equation}
  \label{eq:npsud}
  \Delta(\tilde q; p_{\perp_{\rm max}},0)
  = \Delta_{\rm pert}(\tilde q; p_{\perp_{\rm max}},p_{\perp_0})
  \Delta_{\rm NP}(\tilde q; p_{\perp_0},0)
\end{equation}
We achieve this by extending $\alpha_s(p_{\perp})$ into the non-perturbative region using the following model
\begin{equation}
  \label{eq:asdef}
  \alpha_s(p_{\perp}) =
  \begin{cases}
    \varphi(p_{\perp}), &p_{\perp}<p_{\perp_0},\\
    \alpha_s^{(\rm pert)} (p_{\perp}), &p_{\perp}\geq p_{\perp_0}.
  \end{cases}
  \
\end{equation}
In order to explore the possibility of a reasonable description of experimental data, we
 have studied in a greater detail two simple choices of the non--perturbative function
$\varphi(p_{\perp})$: flat continuation of $\alpha_s(p_{\perp}<p_{\perp_0})$,
  with a constant value $\varphi_0$, $\alpha_s(p_{\perp}<p_{\perp_0}) = \varphi_0$
  and a quadratic interpolation between the two values $\alpha_s(p_{\perp_0})$ and $\varphi_0=\varphi(0)$:
 $$
    \label{eq:asquad}
    \alpha_s(p_{\perp} < p_{\perp_0}) =  \varphi_0 +
    (\alpha_s(p_{\perp_0}) - \varphi_0) {p_{\perp}^2}\slash{p_{\perp_0}^2} \ .
 $$

In both cases our model is determined by two free parameters $p_{\perp_0}$
and $\varphi_0$.
\section{Data sets and fitting results}
In this section we present some new results of the model which were 
obtained after important improvements of Herwig++'s parton shower,
released with version 2.3.1 of the program.
The main change in the program was a fix for a wrongly applied PDF veto in the parton
shower $\bar{q}\rightarrow\bar{q}\ g$ splittings which, by construction 
of our model could have influence on previously presented results \cite{Gieseke:2007ad}. 
Therefore, we have repeated the procedure described in detail in \cite{Gieseke:2007ad} and have fitted the two 
parameters of our model to the Drell-Yan data from three experiments: 
the fixed target $p$--Cu Fermilab E605 \cite{E605data} $\sqrt{s}=38.8$\,GeV, 
CERN ISR $p$--$p$ collisions experiment R209 \cite{R209data}  at $\sqrt{s}=62\,$GeV 
and CDF Tevatron Run I experiment with energies at $\sqrt{s}=1800$ GeV \cite{CDFdata}.
These experiments cover the  whole spectrum of centre mass energy for 
the Drell-Yan process data sets which are interesting for our studies\footnote{There are more data available but all at even
lower CM energies.}.
\subsection{Parton-level study}
\begin{figure}[h]
 \includegraphics[scale=0.32,angle=270.]{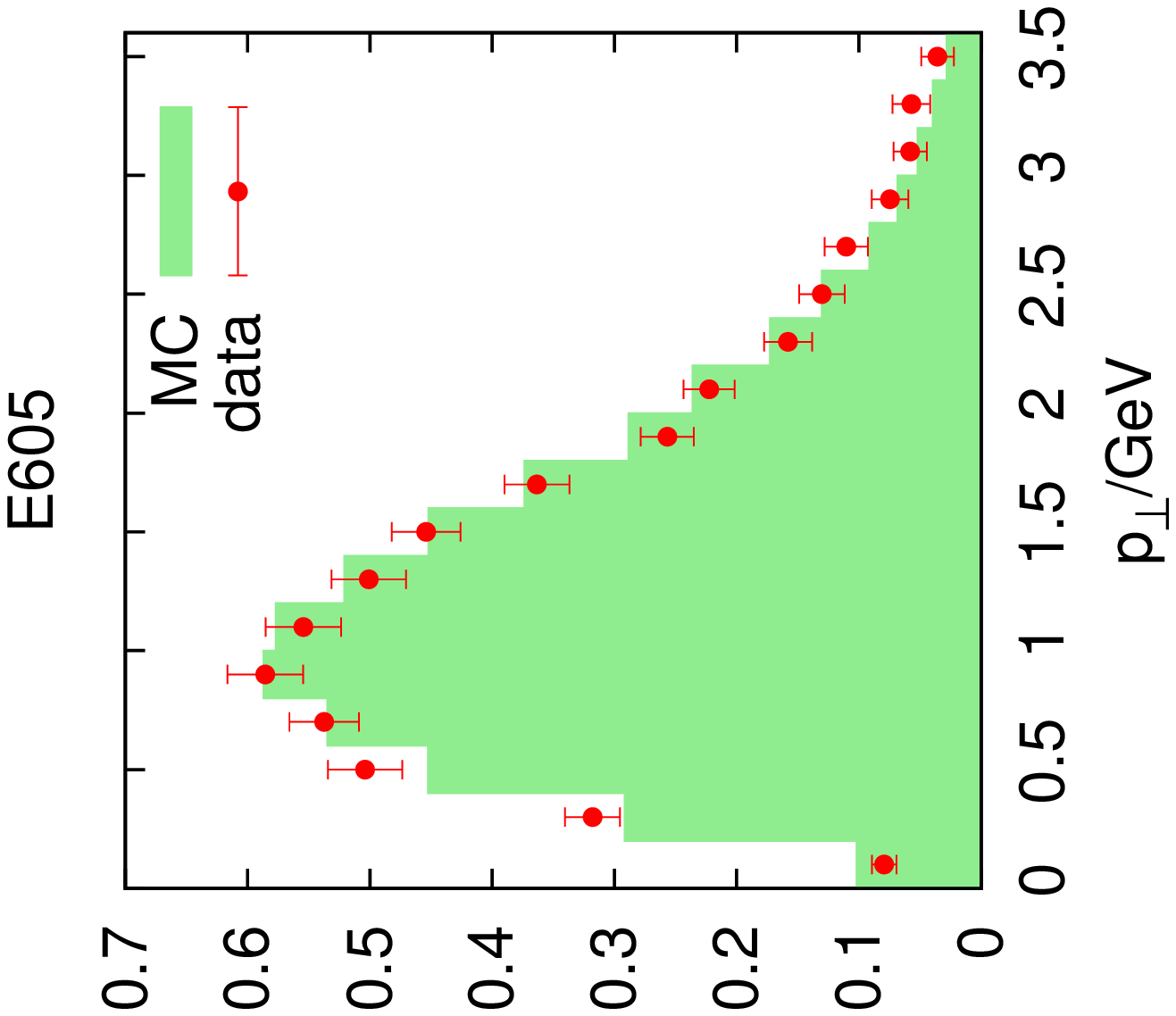}
 \includegraphics[scale=0.32,angle=270.]{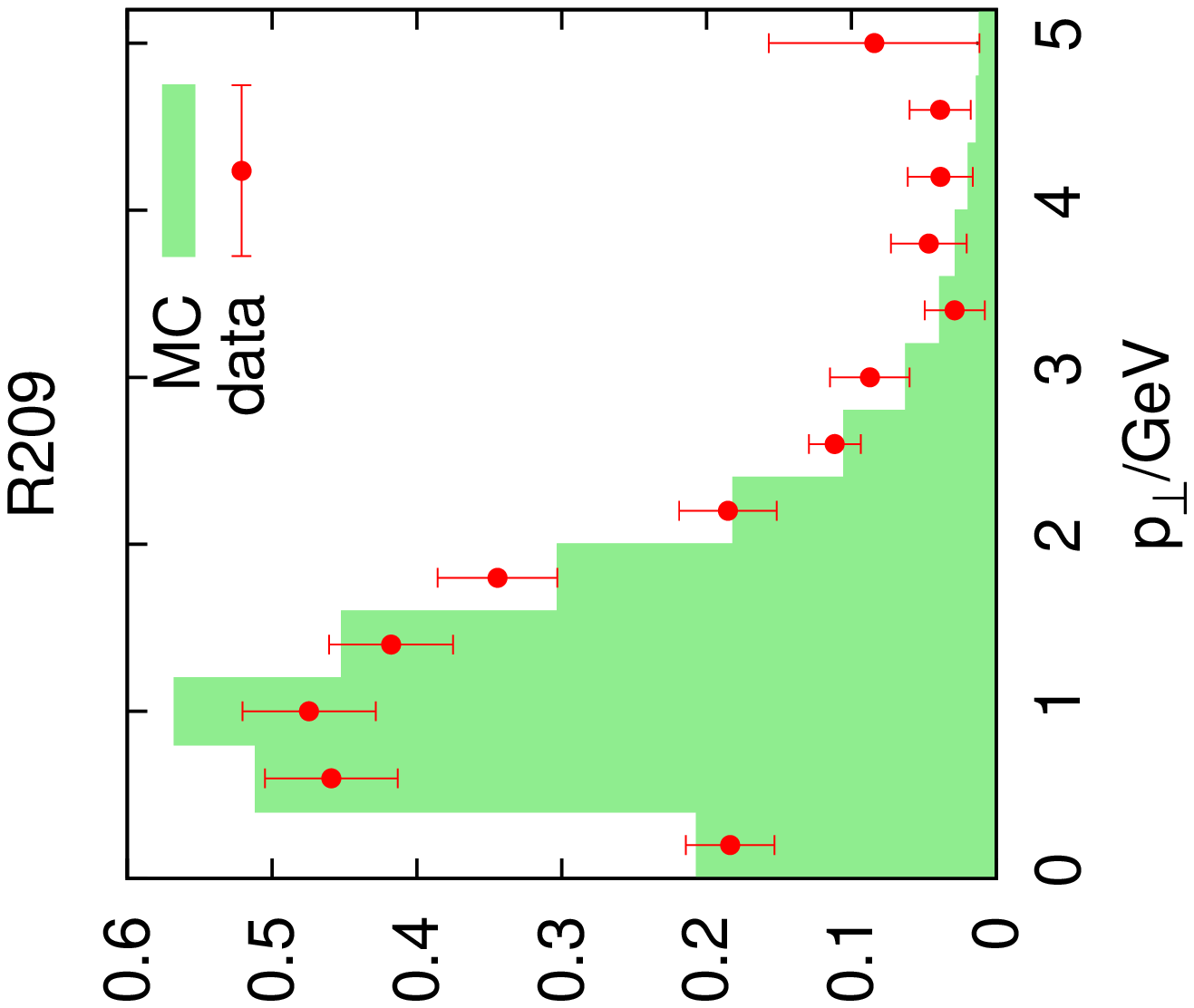}
 \includegraphics[scale=0.32,angle=270.]{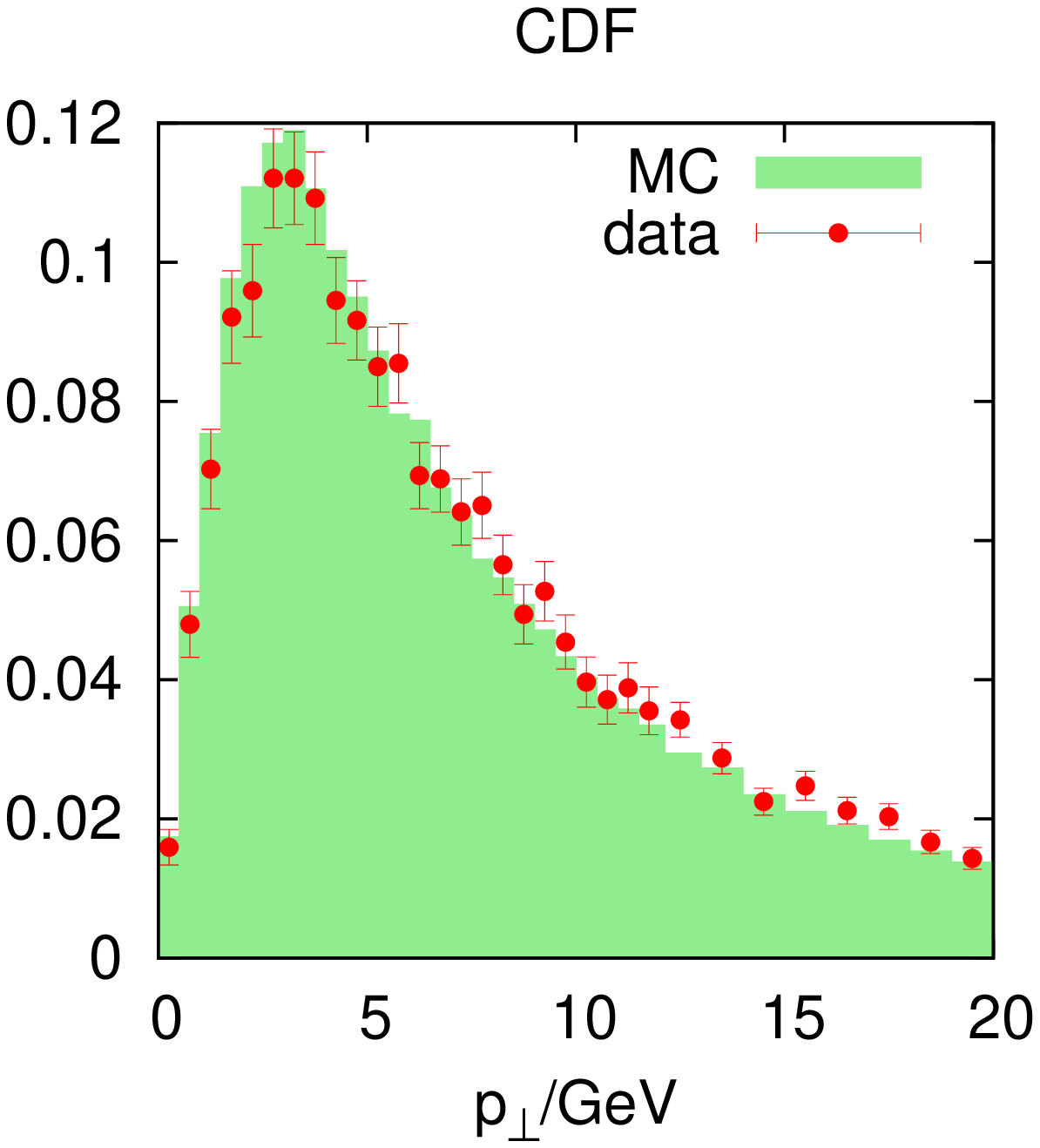}
\caption{The comparisons of the parton-level 
     results from the non--perturbative model
     with the data from E605  with $\chi^2\slash bin = 0.88$ (left), R209 $\chi^2\slash bin = 0.76$ (middle) and CDF $\chi^2\slash bin = 1.0$ (right).
     The Monte Carlo results come from our parameter set with
     $\varphi_0=0.0,\ p_{\perp_0}=0.70\,$GeV. } 
\label{massless}
\end{figure}
In the case of purely parton-level shower, with all the
light-quark and gluon effective masses and cutoffs set to
zero\footnote{For technical reasons, it is not possible to set them
exactly to zero.  However, we have confirmed that if they are small
enough their precise values become irrelevant and have 
very little effect on the results.
},
with our model for the low-scale $\alpha_s$ as the only 
non-perturbative input the fitting procedure gave the 
optimal value for the quadratic extrapolation with $\alpha_s(0)=0.0$ and
 $p_{\perp_0}=0.7$. The resulting low-$p_T$ distributions for the new values 
are presented on top of the data sets in Fig.~\ref{massless}. The $\chi^2\slash bin$ 
values are a little higher than before the parton shower improvements, nevertheless the agreement with data 
 remains at a high-level; 
$\chi^2$ for all the experimental data sets are below or equal one.
If we are only interested in the $W/Z$
transverse momentum distribution, it is enough to use a parton-level study,
however, if one needs to simulate fully exclusive events then a hadronization model
has to be used.
\subsection{Hadron-level results}
The hadronization model used in Herwig++ requires
termination of the shower using non-perturbative effective parton masses tuned to 
$e^+e^-$ data.
\begin{figure}[h]
 \includegraphics[scale=0.32,angle=270.]{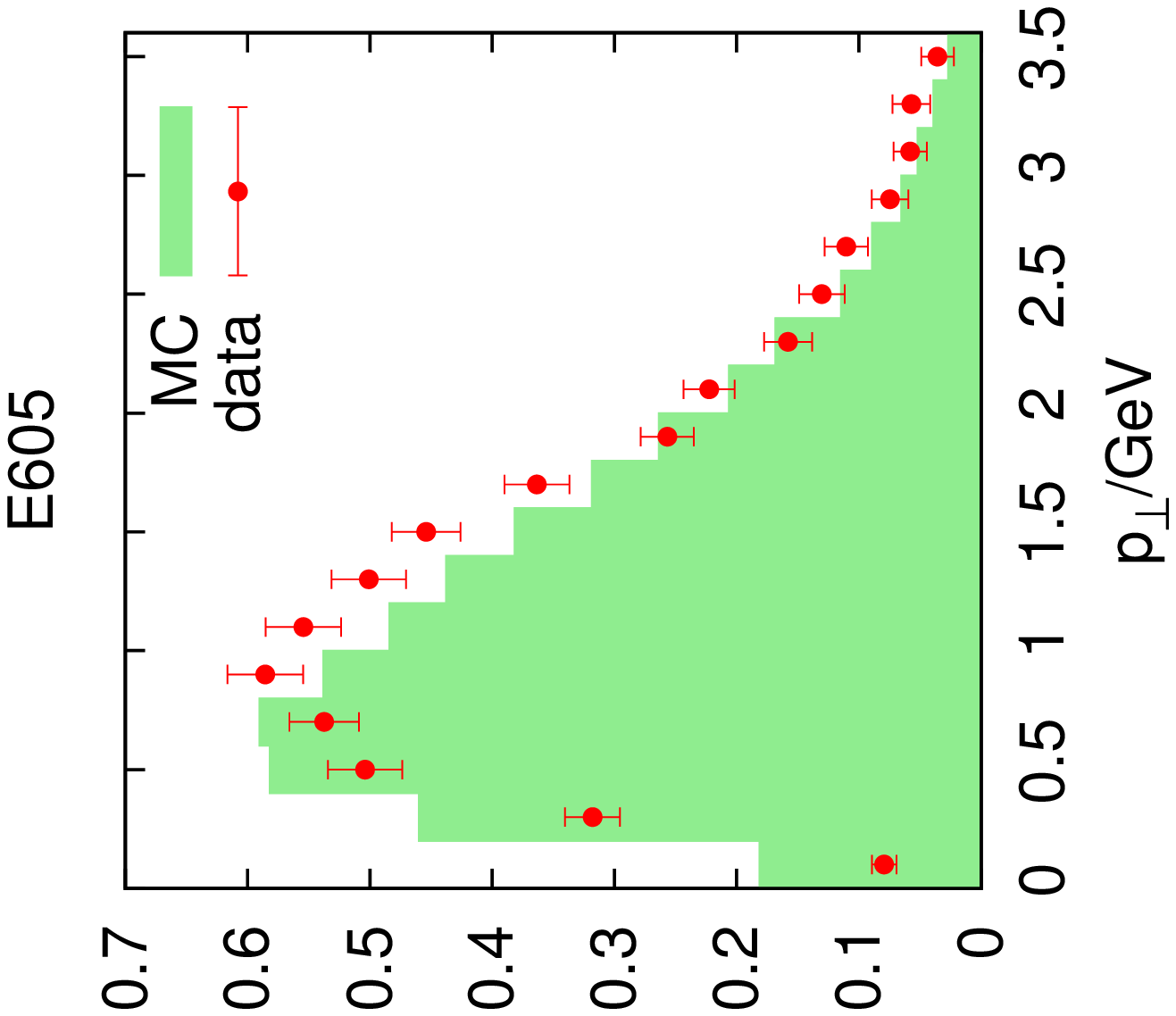}
 \includegraphics[scale=0.32,angle=270.]{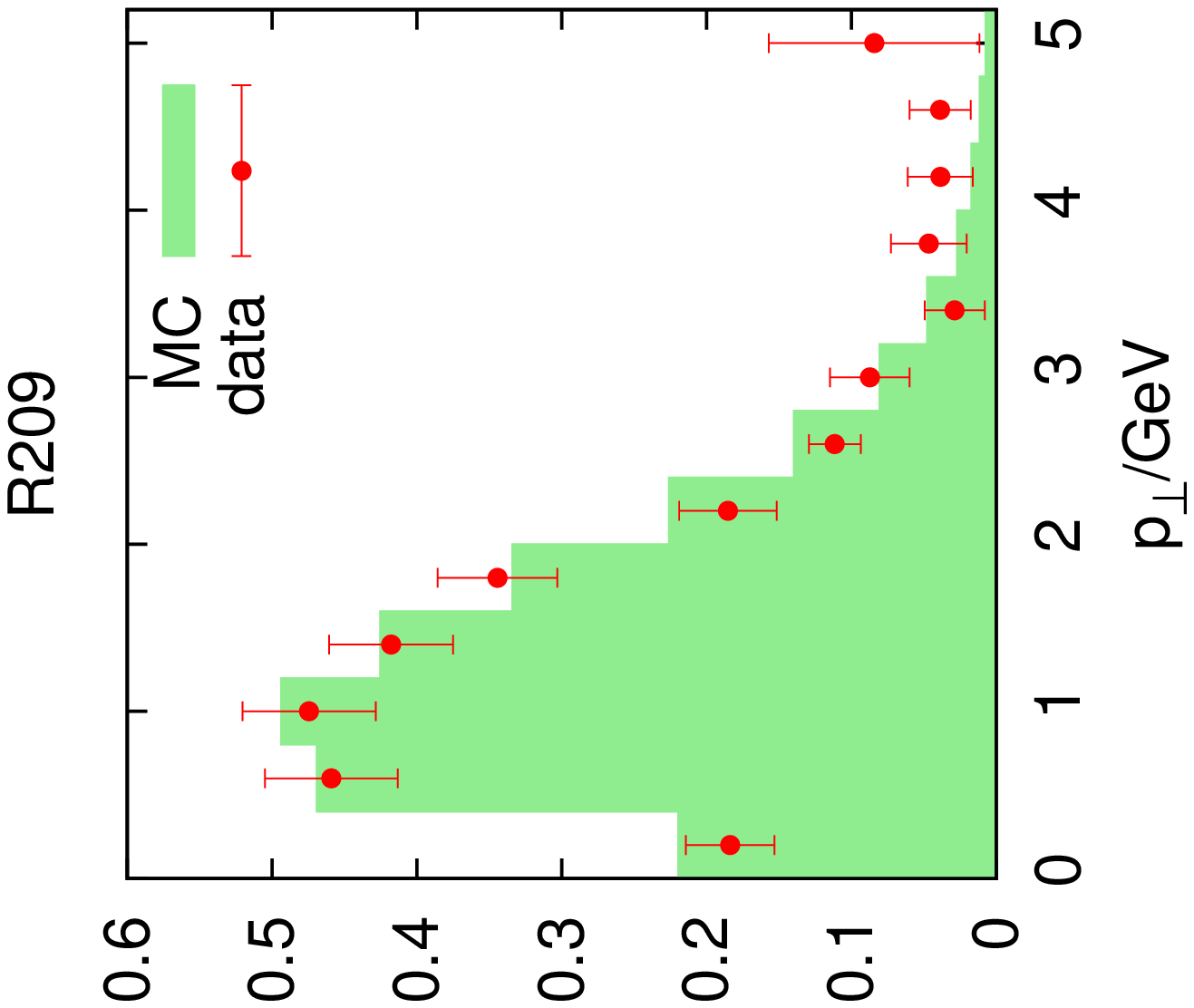}
 \includegraphics[scale=0.32,angle=270.]{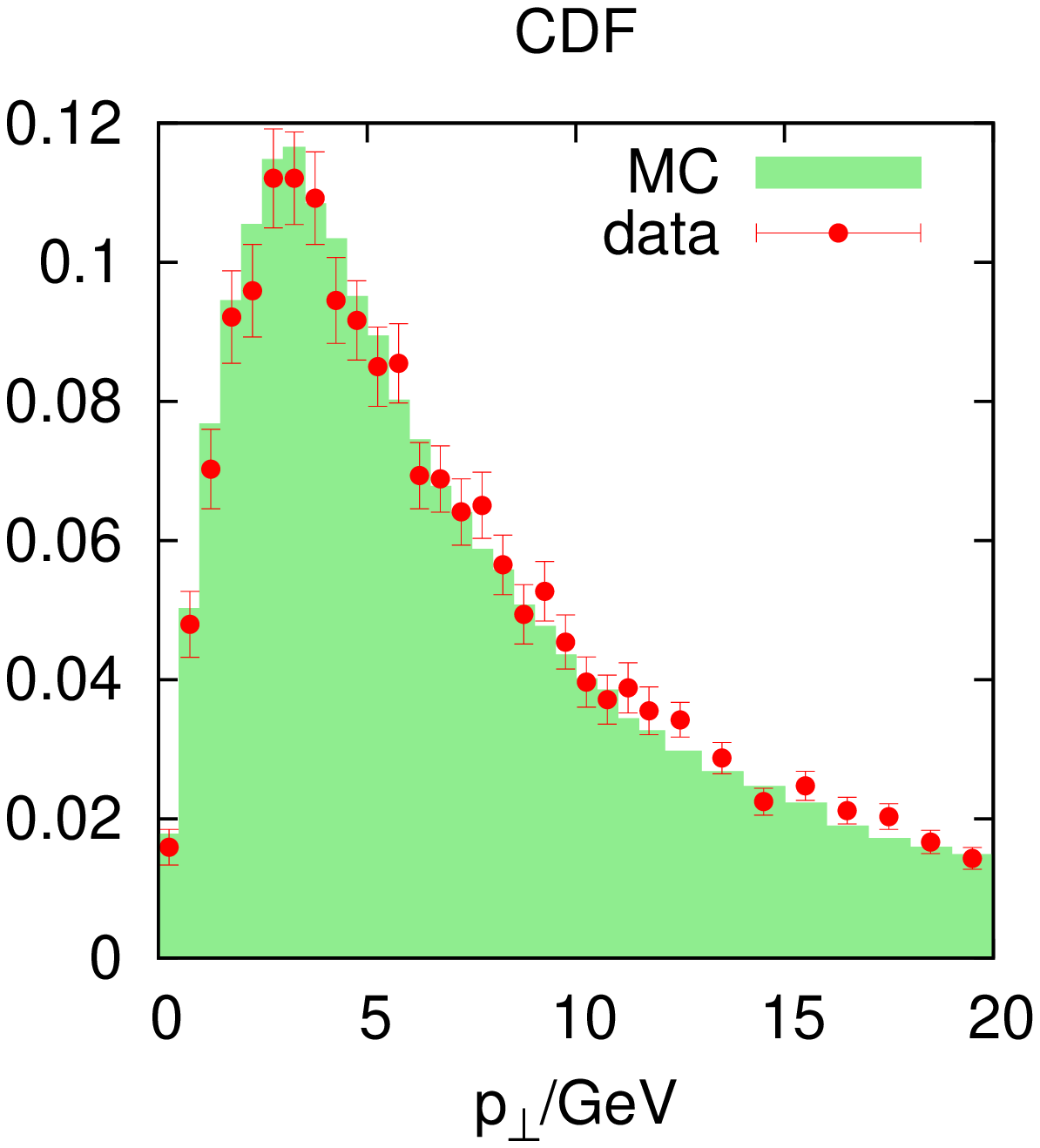}
\caption{Comparison of the hadron level 
     results from the non--perturbative model
     with data from E605  with $\chi^2\slash bin = 8.6$ (left), R209 $\chi^2\slash bin = 0.66$ (middle) and CDF $\chi^2\slash bin = 0.80$ (right).
     The Monte Carlo results are from our parameter set with
     $\varphi_0=4.0, p_{\perp_0}=2.5\,$GeV. }
\label{massive}
\end{figure}
Therefore, we performed the same analysis as above but this time with restored tuned effective parton masses.
In this case the best and most stable situation was found for $\alpha_s(0)=4$ and $p_{\perp_0} = 2.5\,$GeV,
giving the $\chi^2$ per degree of freedom of $0.80$ for CDF, $0.66$ for R209 and the worst $8.6$ for E605.
We should stress that the used parameter set may not be the optimal choice for each experiment
or CM energy but rather the best compromise between the three
experiments.  As the fixed target data in our analysis do not even include the
systematic errors quoted to be around
5--10\%, we have deliberately put a bit more emphasis
on the Tevatron results.  
\subsection{Remarks}
The first remark is that the new parameter choices for both the parton and hadron-level models, 
are not very different from the ones obtained using the old version of Herwig++. Before parton shower improvements our best choices were, for the parton-level mode: $\alpha_s(0)=0.0$, $p_{\perp_0} = 0.75\,$GeV, and for the hadron-level case: $\alpha_s(0)=3$ and $p_{\perp_0} = 3.0\,$GeV  \cite{Gieseke:2007ad}.

We have also checked how the results depend on the intrinsic momentum $k_{\perp}$
by varying its value with $\delta k_{\perp_{\pm}}=\pm 0.1$ GeV around our fixed
value $k_{\perp}=0.4$, which is in the range permitted by the Fermi motion. We have repeated the fitting procedure
and observed that for both intrinsic momenta, $k_{\perp_{\pm}}=k_{\perp}+\delta k_{\perp_{\pm}}$, we are able to find a pair of parameters for which our model gives equally good description of data sets as for the central value of $k_{\perp}=0.4$.
Moreover, we have observed that the value of  $\alpha_s(0)$ parameter for all studied intrinsic momenta remains the same but the $p_{\perp_0}$ value 
is shifted for a bigger intrinsic momentum to 
a higher scale and for a smaller one to a lower scale.
Therefore, by changing the intrinsic momentum from $0.4$ to $0.5$ GeV we can obtain exactly the same best model's parameters set as in  \cite{Gieseke:2007ad} and  the same shape of $\alpha_s$ as presented in Figure 4 from \cite{Gieseke:2007ad}. In that case the 
comparison of the shape of $\alpha_s$ in the non--perturbative region of the parton-level study are in good agreement with other approaches to modelling non-perturbative corrections to inclusive observables with a modified coupling in the soft region \cite{Dokshitzer:1995zt,Dokshitzer:1995qm}.

The last remark is that using our model as the only non-perturbative
ingredient in the simulation, i.e.\ removing the non-perturbative
constituent parton masses that usually cut off the parton shower in
Herwig++, gives a somewhat better description of the data.  This lays
open the speculation that perhaps, in some way, the two approaches
could be combined. One could for example use our model for the initial-state radiation, and
the usual model, tuned to describe the final states of $e^+e^-$
annihilation, for final-state radiation.
\section{Predictions for LHC and comparison with other approaches}
\label{LHC}
At the end of this note we would like to compare the results for a transverse momentum distribution of 
the $Z$ boson at the LHC energies using the nonperturbative gluon emission model and two other approaches:
\ResBos{} \cite{Berge:2004nt} and the Gaussian intrinsic $k_\perp$ extrapolation.
But first let us  compare our prediction of the parton level, marked as the filled
histogram in Fig.~\ref{fig:LHC-pt},  and of the hadron level, dot--dashed blue line. 
Both histograms, as expected, give a consistent extrapolation.
\begin{center}
\hspace*{2cm}\begin{figure}[h]%
  \includegraphics[scale=1.0]{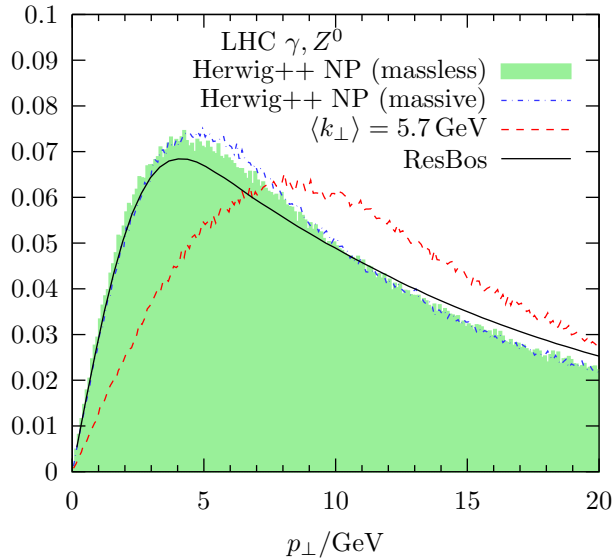}
  \caption{Vector boson $p_\perp$ distribution at the LHC.  Our model
    is compared to the extrapolation of Gaussian intrinsic $k_\perp$
    to LHC energies and the result from \ResBos{}.
    \label{fig:LHC-pt} }
\end{figure}
\end{center}
The result from \ResBos{} in Fig.~\ref{fig:LHC-pt} (solid, black)
shows a slightly different behaviour from our predictions.  We predict
a slightly more prominent peak and a stronger suppression towards
larger transverse momenta.  Both
computations match the data well at large transverse momenta as they
rely on the same hard matrix element contribution for a single hard
gluon emission.  Let us stress the remarkable feature that we 
predict the same peak position with these models which is very important
from the experimental point of view. This feature is quite
understandable as both models are built on the same footing: extra
emissions of soft gluons.  A comparison of \ResBos{} to data from
experiments at various energies including the experiments E605 and R209
was done in \cite{resbos-02}.

Furthermore,  we see the Herwig++ result from only using
intrinsic $\langle k_\perp\rangle = 5.7\,$GeV (dashed, red) as recommended in
\cite{herwigman}\footnote{
This recommendation has been changed and in the latest version of Herwig++ its new value $\langle k_\perp\rangle = 2.2\,$GeV
is adjusted to be in agreement with the prediction of the non-perturbative model presented in Fig.~\ref{fig:LHC-pt}}.
This large value stems from an extrapolation from
lower energy data with the assumption that the average $k_\perp$ will
depend linearly on $\ln(M/\sqrt{s})$.  The peak is seen to lie at a
considerably higher value of the transverse momentum.  It would clearly be
of interest to have experimental data to distinguish these two models
of non--perturbative transverse momentum.

\section{Conclusion}
We consider the model based on soft-gluon radiation,
much like the resummation program \ResBos{}, to have a more meaningful
physics input than simply extrapolating the Gaussian smearing of a
primordial transverse momentum. The model implemented in the 
improved parton shower of \hpp\  (release 2.3.1) 
gives a good (and very similar to 
the older version of \hpp\ ) description of data. 
On the other hand, the fitting procedure shows that the best values for 
the model's parameters are slightly different than the previous ones.
\section{Acknowledgments}
We would like to thank  W. P\l aczek, S. Sapeta and all our colleagues 
from the Herwig++ collaboration for useful discussions.
We gratefully acknowledge support from the EU Marie Curie Research
Training Network MCnet (MRTN-CT-2006-035606) and the Helmholtz Alliance
"Physics at the Terascale". One of us (AS) also would like to 
acknowledge MEN grant N202 175435 (2008-10) for additional support.

\end{document}